\documentclass[a4paper,10pt,aps,prl,twocolumn,superscriptaddress,showpacs,amsmath,amssymb]{revtex4}
\usepackage{graphicx,color}

\newcommand{\dd}{\mathrm{d}}
\newcommand{\td}[2]{\frac{\dd #1}{\dd #2}}

\newcommand{\fd}[2]{\frac{\delta #1}{\delta #2}}

\newcommand{\Int}[1]{\int\dd #1\;}

\renewcommand{\vec}[1]{\mathbf #1}

\newcommand{\al}{\alpha}

\newcommand{\ep}{\epsilon}
\newcommand{\eps}{\varepsilon}
\newcommand{\kap}{\kappa}
\newcommand{\lam}{\lambda}

\newcommand{\sig}{\sigma}

\newcommand{\im}{\text{i}}
\newcommand{\ra}{\rightarrow}
\newcommand{\x}{\vec r}
\newcommand{\Dr}{D_\text{r}}
\newcommand{\vc}{v_\text{c}}
\newcommand{\xc}{\xi_\text{c}}
\newcommand{\qc}{q_\text{c}}
\newcommand{\drho}{\delta\rho}

\begin{document}

\title{Effective Cahn-Hilliard equation for the phase separation of active
  Brownian particles}

\author{Thomas Speck}
\affiliation{Institut f\"ur Physik, Johannes Gutenberg-Universit\"at Mainz,
  Staudingerweg 7-9, 55128 Mainz, Germany}
\author{Julian Bialk\'e}
\author{Andreas M. Menzel}
\author{Hartmut L\"owen}
\affiliation{Institut f\"ur Theoretische Physik II,
  Heinrich-Heine-Universit\"at, D-40225 D\"usseldorf, Germany}

\begin{abstract}
  The kinetic separation of repulsive active Brownian particles into a dense
  and a dilute phase is analyzed using a systematic coarse-graining strategy.
  We derive an effective Cahn-Hilliard equation on large length and time
  scales, which implies that the separation process can be mapped onto that of
  passive particles. A lower density threshold for clustering is found, and
  using our approach we demonstrate that clustering first proceeds via a
  hysteretic nucleation scenario and above a higher threshold changes into a
  spinodal-like instability. Our results are in agreement with
  particle-resolved computer simulations and can be verified in experiments of
  artificial or biological microswimmers.
\end{abstract}


\maketitle


The collective behavior of living ``active'' matter has recently attracted
considerable interest from the statistical physics community (for reviews, see
Refs.~\citenum{rama10,marc13}). Even if the mutual interactions of the
individual units are following simple rules, complex spatiotemporal patterns
can emerge. Examples in nature occur on a wide range of scales from flocks of
birds~\cite{cava09} to bacterial turbulence~\cite{wens12}. A basic physical
model is obtained by describing the individual entities as particles with
internal degrees of freedom (in the simplest case just an orientation) that
consume energy and are thus driven out of thermal equilibrium. Consequently,
shaken granular particles~\cite{nara07} and phoretically propelled colloidal
particles~\cite{pala10,theu12,pala13,butt13} have been investigated in
detail. Moreover, the observed collective behavior might find applications in,
e.g., the sorting~\cite{mija13} and transport of cargo~\cite{pala13a}.

Here we are interested in the phase behavior of \emph{repulsive} particles
below the freezing density. While in equilibrium only one fluid phase exists,
sufficiently dense suspensions of repulsive self-propelled disks undergo an
``active phase separation'', i.e., particles aggregate into a dense,
transiently ordered cluster surrounded by a dilute gas phase. This has been
observed first~\cite{yaou12,redn13} in computer simulations of a minimal
model~\cite{fily13,bial13,sten13}. Clustering has also been reported in
experiments using colloidal suspensions of active Brownian particles, in which
the particles are phoretically propelled along their orientations due to the
catalytic decomposition of hydrogen peroxide on a platinum
hemisphere~\cite{theu12}, or due to light-activated hematite~\cite{pala13}. In
these experiments, phoretic attractive forces play an important role. A closer
realization of ideally repulsive particles is possible through the reversible
local demixing of a near-critical water-lutidine
mixture~\cite{butt12}. Colloidal particles propelled due to the ensuing local
density gradients show indeed the predicted phase
separation~\cite{butt13}. While in passive suspensions phase-separation occurs
only for sufficiently strong attractive forces, the microscopic mechanism for
repulsive active particles is due to self-trapping: colliding particles block
each other due to the persistence of their orientation~\cite{butt13}. In
sufficiently dense suspensions, the ``pressure'' of the free, fast particles
leads to the growth of small clusters until phase separation is reached. This
generic dynamical instability due to a density-dependent mobility has been
first studied by Tailleur and Cates for the run-and-tumble motion of
bacteria~\cite{tail08} and later also for active Brownian
particles~\cite{cate13}. At even higher densities, first steps have been taken
to study glassy dynamics~\cite{ran13,bert13b} and
crystallization~\cite{bial12,menz13}. Another interesting question is the
interplay of the propulsion with attractive
forces~\cite{line12,redn13a,mogn13}.

While the phase separation and nucleation in passive suspensions has been
studied extensively, an open fundamental question is whether the clustering of
active Brownian particles, which is an intrinsically nonequilibrium system,
can be mapped on the phase separation dynamics of passive particles. In this
Letter, we demonstrate for a simple model system of Brownian particles that
this mapping exists on coarse-grained length and time scales. We find a formal
analogy with the Cahn-Hilliard equation, which indeed implies a mapping of the
phase separation of active Brownian particles to the phase separation process
of passive particles governed by attractive forces. This allows to translate
established concepts to the study of active systems and, moreover, implies
that statistical properties (growth exponent, scaling of clusters, etc.) are
unchanged in active systems. We comment that recently additional terms in the
dynamics of phase separation have been studied~\cite{sten13,witt13a}, which
are not derivable from an effective free energy. Our crucial point here is
that simple, genuinely active systems do not generally need these
non-Hamiltonian terms in order to be described correctly. Moreover, we predict
by a weakly nonlinear stability analysis -- and confirm our prediction through
particle-resolved computer simulations -- that the nature of the separation
process (spinodal decomposition or hysteretic nucleation-like behavior)
depends on the propulsion speed along the instability line. While it
corresponds to a spinodal decomposition for small speeds (high density), it
changes to a hysteretic nucleation-like behavior upon crossing a
threshold. The actual instability line predicted by our analysis is in good
agreement with the simulation data.


The minimal model for active Brownian
particles~\cite{yaou12,redn13,fily13,bial13,sten13} that we study consists of
$N$ repulsive disks in two dimensions, the motion of which is governed by
\begin{equation}
  \label{eq:lang}
  \dot\x_k = -\nabla U + v_0\vec e_k + \boldsymbol\eta_k.
\end{equation}
Particles interact via the potential energy $U$, and $\boldsymbol\eta_k$ is the
Gaussian white noise describing the influence of the solvent. In addition,
particles are propelled with constant speed $v_0$ along their orientations
$\vec e_k$, which undergo free rotational diffusion with diffusion coefficient
$\Dr$ and are, therefore, uncorrelated. In an effort to connect these
microscopic equations of motion with the emerging large-scale behavior of the
suspension, we have recently derived the effective hydrodynamic equations
\begin{gather}
  \label{eq:rho}
  \partial_t\rho = -\nabla\cdot[v(\rho)\vec p - D\nabla\rho], \\
  \label{eq:p}
  \partial_t\vec p = -\frac{1}{2}\nabla[v(\rho)\rho] + D\nabla^2\vec p -
  \Dr\vec p
\end{gather}
starting from the full $N$-body Smoluchowski equation for the evolution of the
joint probability distribution of all particle positions and their
orientations~\cite{bial13}. Here, $D$ denotes the long-time diffusion
coefficient of the passive suspension. Eqs.~(\ref{eq:rho}) and (\ref{eq:p})
have been derived under the assumption of spatially slowly varying number
density $\rho(\x,t)$ and orientational field $\vec p(\x,t)$. Instead of
assuming a phenomenological functional form for the effective speed $v(\rho)$
(see Refs.~\cite{cate10,yaou12}), we have shown that for the minimal model
close to the instability line the linear relation $v(\rho)=v_0-\rho\zeta$
follows, where $\zeta$ quantifies the force imbalance due to the
self-trapping.

Before we continue, we simplify the equations through choosing $1/\Dr$ as the
unit of time, $\sqrt{D/\Dr}$ as the unit of length, and we normalize both
fields by the average density, $\rho\mapsto\bar\rho(1+\drho)$ and $\vec
p\mapsto\bar\rho\vec p$. The equations then read
\begin{gather}
  \label{eq:rho:d}
  \partial_t\drho = -\al\nabla\cdot\vec p + \nabla^2\drho +
  4\xi\nabla\cdot(\vec p\drho), \\
  \label{eq:p:d}
  \partial_t\vec p = -\beta\nabla\drho + \nabla^2\vec p - \vec p +
  4\xi\drho\nabla\drho,
\end{gather}
where we have separated the non-linear terms. The dimensionless coefficients
appearing here are defined as
\begin{equation}
  \xi \equiv \frac{\bar\rho\zeta}{v_\ast}, \quad
  \al \equiv 4(v_0/v_\ast - \xi), \quad \beta \equiv 2(v_0/v_\ast - 2\xi)
\end{equation}
with characteristic speed $v_\ast\equiv4\sqrt{D\Dr}$.


Dropping the non-linear terms in Eqs.~\eqref{eq:rho:d} and~\eqref{eq:p:d}, it
is straightforward to investigate the linear stability of the homogeneous
solution $\drho=0$ and $\vec p=0$. Indeed, depending on the values of the
coefficients $\al$ and $\beta$, the homogeneous density profile might become
instable. The dispersion relation
\begin{equation}
  \label{eq:sig}
  \sig(q) = -\frac{1}{2} - q^2 + \frac{1}{2}\sqrt{1-4\al\beta q^2}
  \approx -(1+\al\beta)q^2
\end{equation}
quantifies the growth rate of a perturbation with wave vector $q$. On large
scales (small $q$), the instability occurs whenever $1+\al\beta<0$. From the
condition $1+\al\beta=0$, we determine the value of the dimensionless force
imbalance coefficient
\begin{equation}
  \label{eq:2}
  \xc = \frac{3}{4}(\vc/v_\ast) - \frac{1}{4}\sqrt{(\vc/v_\ast)^2-1}
\end{equation}
at the onset of the instability for a given critical speed
$\vc$~\cite{bial13}. Clearly, $v_\ast$ is the smallest propulsion speed for
which the instability is possible, $\vc\geqslant v_\ast$.


In the linear analysis, a small initial perturbation grows unbounded. Of
course, due to the non-linear terms implying a coupling to other modes, the
amplitude of the perturbation will saturate. We now aim to derive an equation
of motion that describes the evolution of an initial perturbation for
propulsion speeds $v_0=\vc(1+\eps)$ in the vicinity of the linear stability
limit~\cite{cros93}. The fastest growing wave vector following
Eq.~\eqref{eq:sig} is $\qc=\frac{1}{2}\sqrt{(\al\beta)^{-1}-\al\beta}$, which
dominates the initial stage of the developing instability. Expanding
$\al=\al_0+\eps\al_1+\cdots$ and $\beta=\beta_0+\eps\beta_1+\cdots$ we find
$\qc\sim\sqrt\eps$ with $\al_0\beta_0=-1$. The growth rate of this mode is
$\sig(\qc)\approx-\eps\sig_1\qc^2\sim\eps^2$ to lowest order, where we have
defined $\sig_1\equiv\al_0\beta_1+\al_1\beta_0$.

We are interested in the large-scale behavior of the suspension. As suggested
by the scaling of critical wave vector and growth rate, we rescale length with
$1/\sqrt\eps$ and time with $1/\eps^2$, amounting to
$\partial_t\mapsto\eps^2\partial_t$ and
$\nabla\mapsto\sqrt\eps\nabla$. Matching powers suggests to expand
\begin{equation}
  \label{eq:expan}
  \drho = \eps c + \eps^2 c^{(2)} + \cdots, \quad \vec p = \sqrt\eps[\eps\vec
  p^{(1)} + \eps^2\vec p^{(2)} + \cdots].
\end{equation}
To lowest order in $\eps$, we find $\vec p^{(1)}=-\beta_0\nabla c$ for the
orientational field leading to
\begin{equation}
  \label{eq:lin}
  0 = (1+\al_0\beta_0)\nabla^2 c,
\end{equation}
which reproduces the result of the linear stability analysis as
required. Gathering terms of the next order leads to
\begin{gather*}
  \partial_t c = -\al_0\nabla\cdot\vec p^{(2)} - \al_1\nabla\cdot\vec p^{(1)}
  + \nabla^2 c^{(2)} + 4\xc\nabla\cdot[c\vec p^{(1)}], \\
  0 = -\beta_0\nabla c^{(2)} - \beta_1\nabla c + \nabla^2\vec p^{(1)} - \vec
  p^{(2)} + 4\xc c\nabla c.
\end{gather*}
Solving the second equation for $\vec p^{(2)}$ and plugging the result
together with $\vec p^{(1)}$ into the first equation, we first note that the
terms containing $c^{(2)}$ drop out. We, therefore, obtain an evolution
equation for the large-scale density fluctuations $c(\x,t)$ alone,
\begin{equation}
  \label{eq:ch}
  \partial_t c = \sig_1\nabla^2c - \nabla^4c - 2g\nabla\cdot(c\nabla c) =
  \nabla^2\fd{F}{c},
\end{equation}
which is the central result of this Letter. Here,
$g\equiv2\xc(\al_0+\beta_0)\geqslant0$ determines the strength of the
non-linear term, where the equal sign holds for the smallest possible critical
speed $\vc=v_\ast$.

We recognize Eq.~\eqref{eq:ch} as the celebrated Cahn-Hilliard
equation~\cite{cahn58} routinely employed to study phase separation
dynamics. It implies the existence of an effective free energy functional
\begin{equation}
  \label{eq:F}
  F[c] = \Int{\x} \left[\frac{1}{2}|\nabla c|^2 + f(c) \right]
\end{equation}
with bulk free energy density $f(c)=\frac{1}{2}\sig_1 c^2-\frac{1}{3}g
c^3$. Following our analysis, no ``active'' non-integrable terms enter the
interfacial free energy. The expression for $f(c)$ found here misses the
customary $c^4$ term stabilizing the high density phase at a finite value for
the density. Physically, there is an upper bound to $c$ due to the volume
exclusion between particles. This is not contained in the effective
hydrodynamic description~\eqref{eq:rho} and~\eqref{eq:p} of a single tagged
particle but complementarily considered in the following.

\begin{figure}[t]
  \centering
  \includegraphics{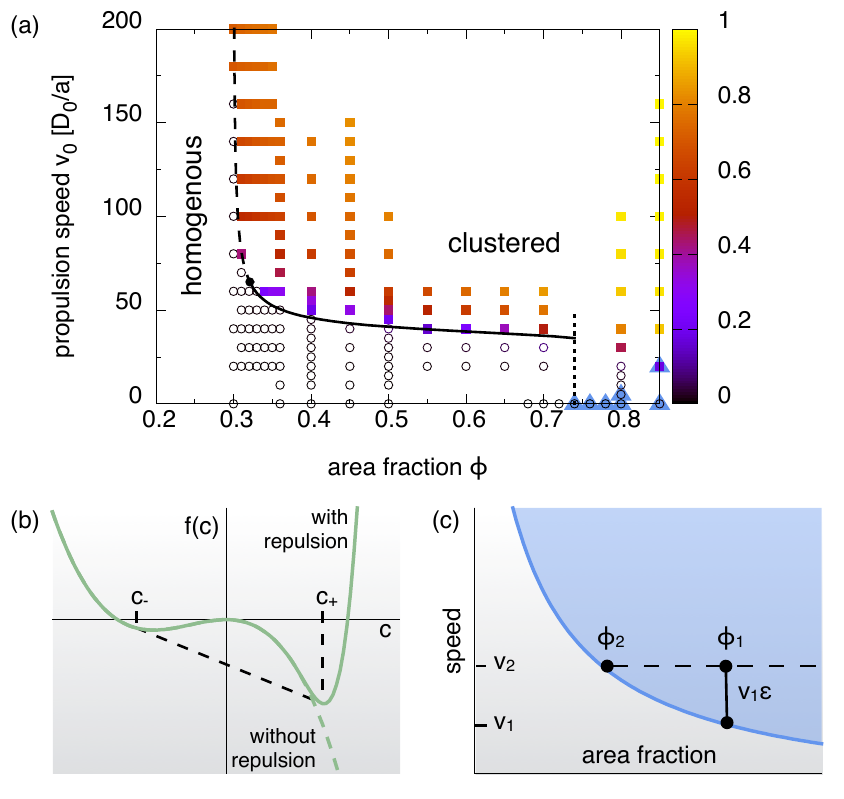}
  \caption{(a)~Instability diagram for repulsive self-propelled disks as a
    function of area fraction $\phi$ and propulsion speed $v_0$. The symbol
    color indicates the fraction of particles that are part of the largest
    cluster. The open symbols correspond to the homogeneous suspension
    ($P<0.1$) and closed squares to the cluster phase. The vertical dotted
    line marks the freezing density for the passive ($v_0=0$) suspension,
    closed triangles indicate solid. The instability line is calculated from
    the analytical result Eq.~\eqref{eq:cc}. A change from a continuous
    (solid) to a discontinuous (dashed) transition at the density
    $\phi_0\simeq0.32$ occurs. (b)~Illustration of the double tangent
    construction. We also sketch the bulk free energy density $f(c)$ for
    $\sig_1<0$ (dashed line). (c)~Derivation of Eq.~\eqref{eq:cc}: Given a
    point $(\phi_1,v_1)$ on the instability line, at the slightly larger speed
    $v_2=v_1(1+\eps)$ phase separation into the dense phase and the dilute gas
    phase with area fraction $\phi_2=\phi_-$ will occur.}
  \label{fig:phasedia}
\end{figure}

We now want to test to what extent such an effective free energy agrees with
particle-resolved simulation data. To this end, we have performed Brownian
dynamics simulations of Eq.~\eqref{eq:lang} for $N=4900$ particles. Particles
interact pairwisely via the repulsive WCA potential, the parameters of which
have been obtained previously by matching experimental data~\cite{butt13}. For
the simulations, we fix the particle diameter $a$, the free diffusion
coefficient $D_0$, and the rotational diffusion coefficient $\Dr=3D_0/a^2$. We
vary the propulsion speed $v_0$ and the area fraction $\phi=N\pi
a^2/(2L)^2=(\pi a^2/4)\bar\rho$, where $L$ is the edge length of the
simulation box employing periodic boundary conditions. We measure the degree
of clustering through the average fraction $P$ of particles that are part of
the largest cluster, cf. Refs.~\cite{butt13,bial13}, which is determined from
steady state trajectories. We equilibrate the passive suspension at the
desired density, turn on $v_0$, and let the system relax into the steady
state. The phase diagram is presented in Fig.~\ref{fig:phasedia}(a), where for
every simulated state point $(\phi,v_0)$ the order parameter $P$ is shown. We
use a simple threshold such that for $P\geqslant0.1$ we consider the
suspension to be in the cluster phase as indicated by a closed symbol. We also
measure the bond orientational order to decide whether the suspension has
become a solid as indicated by triangles. In qualitative agreement with other
simulations~\cite{bial12,fily13}, the propulsion melts the solid before
entering the cluster phase.

In Fig.~\ref{fig:phasedia}(b), the bulk free energy density $f(c)$ is sketched
for speeds $v_0>\vc$ slightly above the critical speed. The form of $f(c)$
implies that the homogeneous density profile with $c=0$ becomes
unstable. Following the double tangent construction, phase separation into a
dense phase with $c_+$ and a dilute gas phase with $c_-$ will occur. The
corresponding area fractions $\phi_\pm=\phi(1+\eps c_\pm)$ follow from the
expansion Eq.~\eqref{eq:expan}. From the simulations, we expect the area
fraction $\phi_+$ of the dense phase to be nearly close-packed.

In order to obtain a more tractable expression, suppose we know a point
$(\phi_1,v_1)$ on the instability line, see
Fig.~\ref{fig:phasedia}(c). Increasing the speed to $v_2=v_1(1+\eps)$, phase
separation is predicted to occur with area fraction $\phi_-=\phi_1[1+\eps
c_-(\phi_1,v_1)]$ of the gas phase. Hence, with $\phi_2=\phi_-$ we have found
a second point on the instability line. Eliminating $\eps$ and taking the
limit $v_2\ra v_1$ leads to the equation
\begin{equation}
  \label{eq:cc}
  \td{\phi}{v} = \frac{\phi}{v}c_-(\phi,v),
\end{equation}
which is formally equivalent to the Clausius-Clapeyron equation quantifying
the slope along the instability line. However, here the system is
intrinsically driven out of equilibrium. We estimate the instability line by
numerically solving Eq.~\eqref{eq:cc}. To this end, we approximate
$c_-(\phi,v)\approx\sig_1/g$ by the local minimum of $f(c)$. While we found an
analytical expression for $g(v_0/v_\ast)$, the coefficient $\sig_1$ is
difficult to estimate from the simulations. As indicated by the numerical
phase diagram Fig.~\ref{fig:phasedia}(a), there is not only a minimal speed
$v_\ast$ but also a minimal density $\phi_\ast$ for clustering to occur. Close
to this lower density $\sig_1(\phi)\propto\phi_\ast-\phi$ should hold and we
use this expression for $\sig_1$ throughout with $\phi_\ast=0.29$ as a fit
parameter. To obtain a continuous function $v_\ast(\phi)$, we fit the
numerically determined long-time diffusion coefficients $D(\phi)$ of the
passive suspension with a quadratic function~[SM]. As demonstrated in
Fig.~\ref{fig:phasedia}(a), despite these approximations we obtain excellent
agreement with the numerical data.


\begin{figure}[t]
  \centering
  \includegraphics{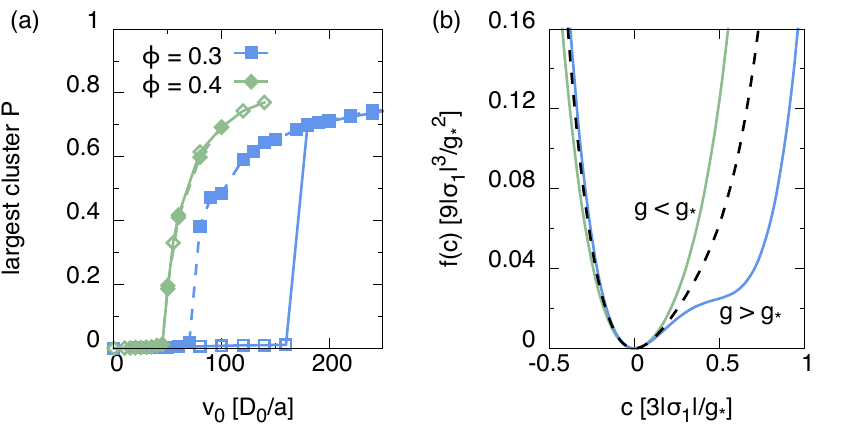}
  \caption{(a)~Hysteresis at low area fraction $\phi$ and vanishing at higher
    values of $\phi$. Shown is the steady-state mean fraction $P$ of particles
    in the largest cluster that is reached for two initial conditions:
    starting from the homogeneous disordered passive suspension (solid lines,
    open symbols) and already containing an ordered cluster (dashed lines,
    closed symbols). (b)~Free energy density ${f}(c)$ for $\sig_1>0$
    containing an additional $c^4$ term that stabilizes the high density
    phase. The dashed line indicates $g=g_\ast$. For $g>g_\ast$, the energy
    density develops a non-convex part indicating coexistence of regions of
    two different particle densities, which promotes hysteresis.}
  \label{fig:hyst}
\end{figure}

A striking observation is made when going to lower densities
$\phi_\ast<\phi\lesssim0.32$, where clustering requires larger propulsion
speeds. Here the order parameter $P$ seems to jump, which is in contrast to
the continuous transition observed at higher densities. To further investigate
this change, we have performed additional simulations, where we prepare a
large ordered cluster consisting of $N/2$ particles as the initial state. We
let the system relax for a finite time ($t_\text{r}=50$) before we record the
data. Fig.~\ref{fig:hyst}(a) shows that at larger densities indeed no
hysteresis is observed, i.e., irrespective of the initial state (disordered or
containing a cluster) the same steady state is reached. This is quite
different for $\phi=0.3$, where a large hysteresis loop can be found. Hence,
we conclude that there is a point ($\phi_0\simeq0.32$ with $\vc\simeq66$) on
the instability line where the transition changes its nature from continuous
to discontinuous. The continuous case is usually described as ``spinodal
decomposition'', whereas the discontinuous behavior of the order parameter
agrees with a nucleation scenario in which a sufficiently large critical
nucleus has to form in order for phase separation to proceed.

Quite remarkably, this change is already contained in the mean-field
description of the Cahn-Hilliard equation (see, e.g., Ref.~\cite{novi85}). For
a qualitative insight, let us discuss the amplitude $|a|$ of ``roll''
perturbations $c(\x)=ae^{\im\vec q\cdot\x}+\text{c.c.}$~[SM]. For $\sig_1<0$,
the non-trivial solution for the amplitude reads
\begin{equation}
  \label{eq:amp}
  |a| \sim \sqrt{-\sig_1/(g_\ast^2-g^2)}.
\end{equation}
This solution exists for $g<g_\ast$ with a threshold $g_\ast\propto q$
proportional to the wave vector $q$ of the destabilizing perturbation. In this
case the bifurcation is supercritical and thus indeed corresponds to a
continuous growth of the amplitude as we push the system deeper into the
instability region. However, when $g$ reaches $g_\ast$, this supercritical
solution ceases to exist.  Rather the transition becomes subcritical, i.e.,
there is a finite region where a stable spatially homogeneous solution $|a|=0$
and a stable solution of non-zero amplitude $|a|\neq0$ coexist and are
separated by an intermediate unstable solution. In Fig.~\ref{fig:hyst}(b), the
bulk free energy density $f(c)$ is plotted for the different regimes, showing
that for $g>g_\ast$ it becomes non-convex, which promotes a discontinuous
course of the transition.


In summary, starting from the effective hydrodynamic equations obtained
previously~\cite{bial13}, we have derived an equation of motion
[Eq.~\eqref{eq:ch}] for the large-scale density fluctuations in a suspension
of active Brownian particles close to the limit of linear stability. This
evolution equation is known from the study of phase separation dynamics in
passive systems as the Cahn-Hilliard equation. In particular, it implies an
effective, although \emph{asymmetric}, free energy without ``non-integrable''
terms, in spite of the genuine activity of the system.  Instead of performing
the double tangent construction explicitly, we have derived Eq.~\eqref{eq:cc}
quantifying the slope of the phase boundary. We have demonstrated excellent
agreement with particle-resolved Brownian dynamics simulations. Moreover,
there is a change of the transition from continuous to discontinuous, which is
also in agreement with the effective theory developed here. An open question
is the exponent for the scaling of the coarsening length of domains. While the
Cahn-Hilliard equation for a conserved order parameter implies the exponent
$1/3$, computer simulations of active Brownian particles in two dimensions
have reported somewhat lower exponents~\cite{redn13,sten13}. However, these
simulations might still be in a transient regime as is also speculated in
Ref.~\cite{witt13a}. Nevertheless, this point will have to be resolved in the
future and calls for further experimental investigations.


We gratefully acknowledge support by the Deutsche Forschungsgemeinschaft
through the recently established priority program 1726.


\section{Strength of the non-linear term}

Here we give the explicit expression for $g=2\xc(\al_0+\beta_0)$. In the main
text, we have defined
\begin{equation}
  \xi \equiv \frac{\bar\rho\zeta}{v_\ast}, \quad
  \al \equiv 4(v_0/v_\ast - \xi), \quad \beta \equiv 2(v_0/v_\ast - 2\xi).
\end{equation}
Moreover, at the linear stability limit, we have
\begin{equation}
  \xc = \frac{3}{4}u - \frac{1}{4}\sqrt{u^2-1}
\end{equation}
with dimensionless critical speed $u\equiv\vc/v_\ast$. At this critical speed
\begin{equation}
  \al_0 = 4(u-\xc), \qquad \beta_0 = 2(u-2\xc).
\end{equation}
We can thus express the non-linear coefficient $g$ in terms of $u$,
\begin{equation}
  \begin{split}
    g(u) &= 2\xc(\al_0+\beta_0) = 4\xc(3u-4\xc) \\ 
    &= 1-u^2+3u\sqrt{u^2-1} \\ &\geqslant 0,
  \end{split}
\end{equation}
where the equal sign holds for $u=1$.

\section{Minimal velocity}

\begin{figure}[h]
  \centering
  \includegraphics{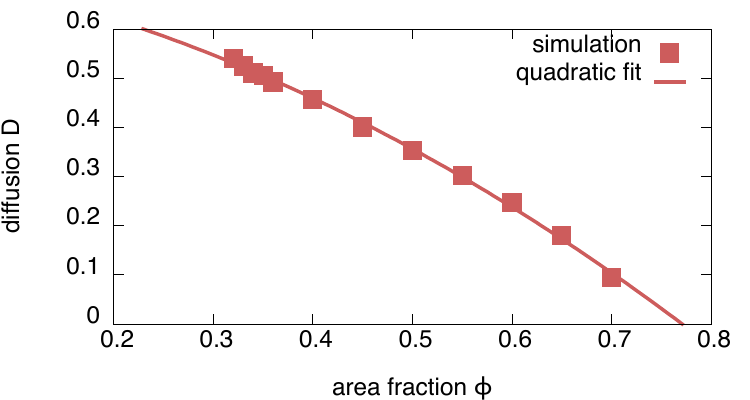}
  \caption{Long-time diffusion coefficients of the passive ($v_0=0$)
    suspension as a function of the area fraction $\phi$.}
  \label{fig:diff}
\end{figure}

The minimal critical velocity for the dynamical phase separation to be
possible is $v_\ast=4\sqrt{D\Dr}$, where we estimate $D$ from the long-time
passive diffusion coefficient. The numerically determined diffusion
coefficients are well fitted by the quadratic function
\begin{equation}
  D(\phi) \simeq -0.76\phi^2 - 0.34\phi + 0.72,
\end{equation}
see Fig.~\ref{fig:diff}.

\section{Amplitude equation}

\begin{figure}[t]
  \centering
  \includegraphics{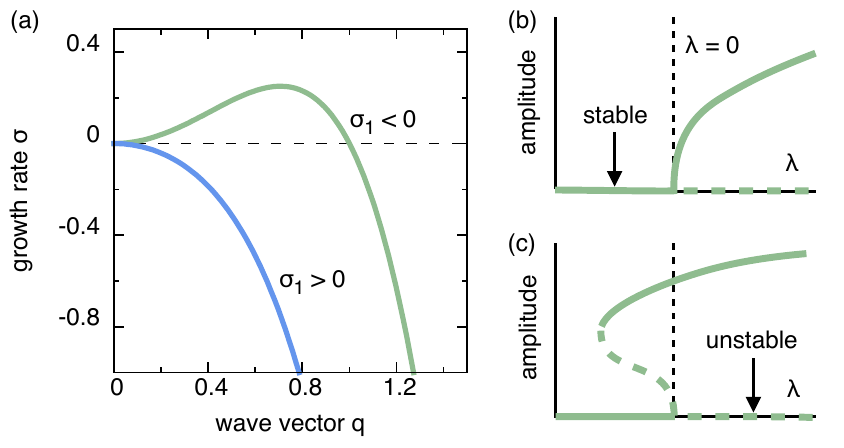}
  \caption{(a)~Growth rate as a function of the wave vector $q$ for the
    linearly unstable ($\sig_1<0$) and linearly stable ($\sig_1>0$)
    regime. (b)~Bifurcation diagram for the supercritical (continuous
    transition) and (c)~the subcritical (discontinuous transition) case.}
  \label{fig:bifur}
\end{figure}

For completeness, we present the derivation of the amplitude equation for roll
solutions following standard arguments, see Ref.~\cite{cros93} for a
comprehensive review. Specifically, we follow the route described in
Ref.~\cite{novi85}. Since we are mainly interested in the qualitative
behavior, we simply augment the effective free energy with a term proportional
to $c^4$ in order to take into account the volume exclusion. The Cahn-Hilliard
equation then reads
\begin{equation}
  \label{eq:2}
  \partial_t c = \nabla^2\left[\sig_1c - g c^2 + \kap c^3 - \nabla^2
    c\right].
\end{equation}
From this equation we obtain the linear dispersion relation
$\hat\sig(q)=-\sig_1 q^2-q^4$, which reaches its maximum at $\qc^2=-\sig_1/2$
and becomes zero for $q_0^2=-\sig_1$ in the case of $\sigma_1<0$, see
Fig.~\ref{fig:bifur}(a). The wave vector $q_0$ marks the boundary of linear
stability, i.e.\ for wave vectors $q>q_0$, the unperturbed solution is still
linearly stable.

We aim to investigate periodic structures with a wave vector $q$ close to
$q_0$, $q^2=q_0^2-\lam\ep^2$, where $\lam$ corresponds to the ``quench
depth''. The growth rate is $\hat\sig(q)\sim\lam q_0^2\ep^2$. This suggests to
introduce an even slower time scale $s=\ep^2t$ leading to
\begin{equation}
  \label{eq:c}
  \ep^2\partial_s c = Lc + \nabla^2\left[-\lam\ep^2c - gc^2 + \kap c^3\right]
\end{equation}
with the linear self-adjoint operator $L=-q^2\nabla^2-\nabla^4$ and
$\sig_1=-q^2-\lam\ep^2$. The next step is to expand $c(\x,s)=\ep
c_1(\x,s)+\ep^2c_2(\x,s)+\cdots$ into powers of $\ep$.

Specifically, we are interested in periodic ``roll'' solutions of the form
\begin{equation}
  c_1(\x,s) = a(s)e^{\im\vec q\cdot\x} + \text{c.c.}
\end{equation}
with complex amplitude $a$. To linear order, we find $Lc_1=0$. Hence, the
amplitude of the linear solution would grow in an unbounded way.  However, the
non-linear terms in Eq.~\eqref{eq:c} couple this solution to higher
wave-vectors and thus lead to a saturation of the amplitude. To second order,
we obtain
\begin{equation}
  0 = Lc_2 - g\nabla^2c_1^2.
\end{equation}
We make the ansatz
\begin{equation}
  c_2(\x,s) = b(s)e^{2\im\vec q\cdot\x} + \text{c.c.}
\end{equation}
leading to
\begin{equation}
  0 = -12q^4b + 4gq^2 a^2, \qquad b = \frac{g}{3q^2}a^2.
\end{equation}
To order $\ep^3$, we obtain
\begin{equation}
  \partial_s c_1 = Lc_3 + \nabla^2\left[-\lam c_1 - 2g(c_1c_2) + \kap
    c_1^3\right].
\end{equation}
We multiply this equation by $e^{-\im\vec q\cdot\x}$ and integrate over an
area $A$, where the edge length in the direction of $\vec{q}$ is set to a
multiple of $2\pi/q$. Only terms that have a spatial dependence $e^{\im\vec
  q\cdot\x}$ will survive the integration. Hence, the amplitude equation
becomes
\begin{equation}
  \begin{split}
    \dot a &= q^2\lam a + 2gq^2 a^\ast b - 3\kap q^2 a^\ast a^2 \\
    &= q^2\lam a + \frac{2}{3}\left(g^2 - \frac{9}{2}\kap q^2\right)|a|^2a,
  \end{split}
\end{equation}
where $a^\ast$ denotes the complex conjugate of $a$ and 
$|a|^2=aa^\ast$. The
stationary solutions ($\dot a=0$) are $|a|=0$ and
\begin{equation}
  |a|^2 = \frac{\frac{3}{2}q^2\lam}{\frac{9}{2}\kap q^2-g^2} =
  \frac{\lam}{3\kap}\frac{1}{1-(g/g_\ast)^2}
\end{equation}
with $g_\ast^2=\frac{9}{2}\kap q^2$.

We can distinguish between the two cases $g<g_\ast$ and $g>g_\ast$: (i)~For
$g<g_\ast$ and $\lam<0$ we only find the trivial solution $a=0$. Increasing
$\lam$ to $\lam>0$, the trivial solution becomes unstable and we obtain the
amplitude $|a|\sim\sqrt\lam$. This behavior marks a supercritical bifurcation
corresponding to a continuous transition.  (ii)~On the contrary, for
$g>g_\ast$ only the trivial solution $|a|=0$ follows for $\lam>0$, and it is
unstable. However, two solutions are obtained from the above procedure for
$\lam<0$: $|a|=0$, which is now stable, and an unstable solution
$|a|\sim\sqrt{-\lam}$. Hence, at the threshold $g=g_\ast$, the nature of the
bifurcation changes from supercritical to subcritical. In
Fig.~\ref{fig:bifur}(b) and (c), corresponding bifurcation diagrams are
sketched.

Finally, we rescale $c=(3|\sig_1|/g_\ast)\hat c$ and eliminate $\kap$ in favor
of $g_\ast$. We test for wave vectors of magnitude $q^2=|\sigma_1|/2$, which
corresponds to $q=\qc$ of the fastest growing perturbation in the case of
$\sigma_1<0$. This implies $\lambda=-\sigma_1/2$, and we can write the bulk
free energy density as
\begin{equation}
  f(\hat c) = \frac{9|\sig_1|^3}{g_\ast^2} \left\{ \pm\frac{1}{2}\hat c^2 -
    \frac{g}{g_\ast} \hat c^3 + \hat c^4 \right\}
\end{equation}
for $\sigma_1>0$ and $\sigma_1<0$, respectively.  This function is plotted in
Fig.~2(b) in the main text. In agreement with the analysis of the amplitude
equation, the bulk free energy becomes non-convex for $g>g_\ast$, in line with
a subcritical bifurcation.


\begin{thebibliography}{31}
\expandafter\ifx\csname natexlab\endcsname\relax\def\natexlab#1{#1}\fi
\expandafter\ifx\csname bibnamefont\endcsname\relax
  \def\bibnamefont#1{#1}\fi
\expandafter\ifx\csname bibfnamefont\endcsname\relax
  \def\bibfnamefont#1{#1}\fi
\expandafter\ifx\csname citenamefont\endcsname\relax
  \def\citenamefont#1{#1}\fi
\expandafter\ifx\csname url\endcsname\relax
  \def\url#1{\texttt{#1}}\fi
\expandafter\ifx\csname urlprefix\endcsname\relax\def\urlprefix{URL }\fi
\providecommand{\bibinfo}[2]{#2}
\providecommand{\eprint}[2][]{\url{#2}}

\bibitem[{\citenamefont{Ramaswamy}(2010)}]{rama10}
\bibinfo{author}{\bibfnamefont{S.}~\bibnamefont{Ramaswamy}},
  \bibinfo{journal}{Annu. Rev. Cond. Mat. Phys.} \textbf{\bibinfo{volume}{1}},
  \bibinfo{pages}{323} (\bibinfo{year}{2010}).

\bibitem[{\citenamefont{Marchetti et~al.}(2013)\citenamefont{Marchetti, Joanny,
  Ramaswamy, Liverpool, Prost, Rao, and Simha}}]{marc13}
\bibinfo{author}{\bibfnamefont{M.~C.} \bibnamefont{Marchetti}},
  \bibinfo{author}{\bibfnamefont{J.~F.} \bibnamefont{Joanny}},
  \bibinfo{author}{\bibfnamefont{S.}~\bibnamefont{Ramaswamy}},
  \bibinfo{author}{\bibfnamefont{T.~B.} \bibnamefont{Liverpool}},
  \bibinfo{author}{\bibfnamefont{J.}~\bibnamefont{Prost}},
  \bibinfo{author}{\bibfnamefont{M.}~\bibnamefont{Rao}}, \bibnamefont{and}
  \bibinfo{author}{\bibfnamefont{R.~A.} \bibnamefont{Simha}},
  \bibinfo{journal}{Rev. Mod. Phys.} \textbf{\bibinfo{volume}{85}},
  \bibinfo{pages}{1143} (\bibinfo{year}{2013}).

\bibitem[{\citenamefont{Cavagna}(2009)}]{cava09}
\bibinfo{author}{\bibfnamefont{A.}~\bibnamefont{Cavagna}},
  \bibinfo{journal}{Phys. Rep.} \textbf{\bibinfo{volume}{476}},
  \bibinfo{pages}{51} (\bibinfo{year}{2009}).

\bibitem[{\citenamefont{Wensink et~al.}(2012)\citenamefont{Wensink, Dunkel,
  Heidenreich, Drescher, Goldstein, L\"owen, and Yeomans}}]{wens12}
\bibinfo{author}{\bibfnamefont{H.~H.} \bibnamefont{Wensink}},
  \bibinfo{author}{\bibfnamefont{J.}~\bibnamefont{Dunkel}},
  \bibinfo{author}{\bibfnamefont{S.}~\bibnamefont{Heidenreich}},
  \bibinfo{author}{\bibfnamefont{K.}~\bibnamefont{Drescher}},
  \bibinfo{author}{\bibfnamefont{R.~E.} \bibnamefont{Goldstein}},
  \bibinfo{author}{\bibfnamefont{H.}~\bibnamefont{L\"owen}}, \bibnamefont{and}
  \bibinfo{author}{\bibfnamefont{J.~M.} \bibnamefont{Yeomans}},
  \bibinfo{journal}{Proc. Natl. Acad. Sci. U.S.A.}
  \textbf{\bibinfo{volume}{109}}, \bibinfo{pages}{14308}
  (\bibinfo{year}{2012}).

\bibitem[{\citenamefont{Narayan et~al.}(2007)\citenamefont{Narayan, Ramaswamy,
  and Menon}}]{nara07}
\bibinfo{author}{\bibfnamefont{V.}~\bibnamefont{Narayan}},
  \bibinfo{author}{\bibfnamefont{S.}~\bibnamefont{Ramaswamy}},
  \bibnamefont{and} \bibinfo{author}{\bibfnamefont{N.}~\bibnamefont{Menon}},
  \bibinfo{journal}{Science} \textbf{\bibinfo{volume}{317}},
  \bibinfo{pages}{105} (\bibinfo{year}{2007}).

\bibitem[{\citenamefont{Palacci et~al.}(2010)\citenamefont{Palacci,
  Cottin-Bizonne, Ybert, and Bocquet}}]{pala10}
\bibinfo{author}{\bibfnamefont{J.}~\bibnamefont{Palacci}},
  \bibinfo{author}{\bibfnamefont{C.}~\bibnamefont{Cottin-Bizonne}},
  \bibinfo{author}{\bibfnamefont{C.}~\bibnamefont{Ybert}}, \bibnamefont{and}
  \bibinfo{author}{\bibfnamefont{L.}~\bibnamefont{Bocquet}},
  \bibinfo{journal}{Phys. Rev. Lett.} \textbf{\bibinfo{volume}{105}},
  \bibinfo{pages}{088304} (\bibinfo{year}{2010}).

\bibitem[{\citenamefont{Theurkauff et~al.}(2012)\citenamefont{Theurkauff,
  Cottin-Bizonne, Palacci, Ybert, and Bocquet}}]{theu12}
\bibinfo{author}{\bibfnamefont{I.}~\bibnamefont{Theurkauff}},
  \bibinfo{author}{\bibfnamefont{C.}~\bibnamefont{Cottin-Bizonne}},
  \bibinfo{author}{\bibfnamefont{J.}~\bibnamefont{Palacci}},
  \bibinfo{author}{\bibfnamefont{C.}~\bibnamefont{Ybert}}, \bibnamefont{and}
  \bibinfo{author}{\bibfnamefont{L.}~\bibnamefont{Bocquet}},
  \bibinfo{journal}{Phys. Rev. Lett.} \textbf{\bibinfo{volume}{108}},
  \bibinfo{pages}{268303} (\bibinfo{year}{2012}).

\bibitem[{\citenamefont{Palacci
  et~al.}(2013{\natexlab{a}})\citenamefont{Palacci, Sacanna, Steinberg, Pine,
  and Chaikin}}]{pala13}
\bibinfo{author}{\bibfnamefont{J.}~\bibnamefont{Palacci}},
  \bibinfo{author}{\bibfnamefont{S.}~\bibnamefont{Sacanna}},
  \bibinfo{author}{\bibfnamefont{A.~P.} \bibnamefont{Steinberg}},
  \bibinfo{author}{\bibfnamefont{D.~J.} \bibnamefont{Pine}}, \bibnamefont{and}
  \bibinfo{author}{\bibfnamefont{P.~M.} \bibnamefont{Chaikin}},
  \bibinfo{journal}{Science} \textbf{\bibinfo{volume}{339}},
  \bibinfo{pages}{936} (\bibinfo{year}{2013}{\natexlab{a}}).

\bibitem[{\citenamefont{Buttinoni et~al.}(2013)\citenamefont{Buttinoni,
  Bialk\'e, K\"ummel, L\"owen, Bechinger, and Speck}}]{butt13}
\bibinfo{author}{\bibfnamefont{I.}~\bibnamefont{Buttinoni}},
  \bibinfo{author}{\bibfnamefont{J.}~\bibnamefont{Bialk\'e}},
  \bibinfo{author}{\bibfnamefont{F.}~\bibnamefont{K\"ummel}},
  \bibinfo{author}{\bibfnamefont{H.}~\bibnamefont{L\"owen}},
  \bibinfo{author}{\bibfnamefont{C.}~\bibnamefont{Bechinger}},
  \bibnamefont{and} \bibinfo{author}{\bibfnamefont{T.}~\bibnamefont{Speck}},
  \bibinfo{journal}{Phys. Rev. Lett.} \textbf{\bibinfo{volume}{110}},
  \bibinfo{pages}{238301} (\bibinfo{year}{2013}).

\bibitem[{\citenamefont{Mijalkov and Volpe}(2013)}]{mija13}
\bibinfo{author}{\bibfnamefont{M.}~\bibnamefont{Mijalkov}} \bibnamefont{and}
  \bibinfo{author}{\bibfnamefont{G.}~\bibnamefont{Volpe}},
  \bibinfo{journal}{Soft Matter} \textbf{\bibinfo{volume}{9}},
  \bibinfo{pages}{6376} (\bibinfo{year}{2013}).

\bibitem[{\citenamefont{Palacci
  et~al.}(2013{\natexlab{b}})\citenamefont{Palacci, Sacanna, Vatchinsky,
  Chaikin, and Pine}}]{pala13a}
\bibinfo{author}{\bibfnamefont{J.}~\bibnamefont{Palacci}},
  \bibinfo{author}{\bibfnamefont{S.}~\bibnamefont{Sacanna}},
  \bibinfo{author}{\bibfnamefont{A.}~\bibnamefont{Vatchinsky}},
  \bibinfo{author}{\bibfnamefont{P.~M.} \bibnamefont{Chaikin}},
  \bibnamefont{and} \bibinfo{author}{\bibfnamefont{D.~J.} \bibnamefont{Pine}},
  \bibinfo{journal}{J. Am. Chem. Soc.} \textbf{\bibinfo{volume}{135}},
  \bibinfo{pages}{15978} (\bibinfo{year}{2013}{\natexlab{b}}).

\bibitem[{\citenamefont{Fily and Marchetti}(2012)}]{yaou12}
\bibinfo{author}{\bibfnamefont{Y.}~\bibnamefont{Fily}} \bibnamefont{and}
  \bibinfo{author}{\bibfnamefont{M.~C.} \bibnamefont{Marchetti}},
  \bibinfo{journal}{Phys. Rev. Lett.} \textbf{\bibinfo{volume}{108}},
  \bibinfo{pages}{235702} (\bibinfo{year}{2012}).

\bibitem[{\citenamefont{Redner et~al.}(2013{\natexlab{a}})\citenamefont{Redner,
  Hagan, and Baskaran}}]{redn13}
\bibinfo{author}{\bibfnamefont{G.~S.} \bibnamefont{Redner}},
  \bibinfo{author}{\bibfnamefont{M.~F.} \bibnamefont{Hagan}}, \bibnamefont{and}
  \bibinfo{author}{\bibfnamefont{A.}~\bibnamefont{Baskaran}},
  \bibinfo{journal}{Phys. Rev. Lett.} \textbf{\bibinfo{volume}{110}},
  \bibinfo{pages}{055701} (\bibinfo{year}{2013}{\natexlab{a}}).

\bibitem[{\citenamefont{Fily et~al.}(2013)\citenamefont{Fily, Henkes, and
  Marchetti}}]{fily13}
\bibinfo{author}{\bibfnamefont{Y.}~\bibnamefont{Fily}},
  \bibinfo{author}{\bibfnamefont{S.}~\bibnamefont{Henkes}}, \bibnamefont{and}
  \bibinfo{author}{\bibfnamefont{M.~C.} \bibnamefont{Marchetti}},
  \bibinfo{journal}{arXiv:1309.3714}  (\bibinfo{year}{2013}).

\bibitem[{\citenamefont{Bialk\'e et~al.}(2013)\citenamefont{Bialk\'e, L\"owen,
  and Speck}}]{bial13}
\bibinfo{author}{\bibfnamefont{J.}~\bibnamefont{Bialk\'e}},
  \bibinfo{author}{\bibfnamefont{H.}~\bibnamefont{L\"owen}}, \bibnamefont{and}
  \bibinfo{author}{\bibfnamefont{T.}~\bibnamefont{Speck}},
  \bibinfo{journal}{EPL} \textbf{\bibinfo{volume}{103}}, \bibinfo{pages}{30008}
  (\bibinfo{year}{2013}).

\bibitem[{\citenamefont{Stenhammar et~al.}(2013)\citenamefont{Stenhammar,
  Tiribocchi, Allen, Marenduzzo, and Cates}}]{sten13}
\bibinfo{author}{\bibfnamefont{J.}~\bibnamefont{Stenhammar}},
  \bibinfo{author}{\bibfnamefont{A.}~\bibnamefont{Tiribocchi}},
  \bibinfo{author}{\bibfnamefont{R.~J.} \bibnamefont{Allen}},
  \bibinfo{author}{\bibfnamefont{D.}~\bibnamefont{Marenduzzo}},
  \bibnamefont{and} \bibinfo{author}{\bibfnamefont{M.~E.} \bibnamefont{Cates}},
  \bibinfo{journal}{Phys. Rev. Lett.} \textbf{\bibinfo{volume}{111}},
  \bibinfo{pages}{145702} (\bibinfo{year}{2013}).

\bibitem[{\citenamefont{Buttinoni et~al.}(2012)\citenamefont{Buttinoni, Volpe,
  K\"ummel, Volpe, and Bechinger}}]{butt12}
\bibinfo{author}{\bibfnamefont{I.}~\bibnamefont{Buttinoni}},
  \bibinfo{author}{\bibfnamefont{G.}~\bibnamefont{Volpe}},
  \bibinfo{author}{\bibfnamefont{F.}~\bibnamefont{K\"ummel}},
  \bibinfo{author}{\bibfnamefont{G.}~\bibnamefont{Volpe}}, \bibnamefont{and}
  \bibinfo{author}{\bibfnamefont{C.}~\bibnamefont{Bechinger}},
  \bibinfo{journal}{J. Phys.: Cond. Matter} \textbf{\bibinfo{volume}{24}},
  \bibinfo{pages}{284129} (\bibinfo{year}{2012}).

\bibitem[{\citenamefont{Tailleur and Cates}(2008)}]{tail08}
\bibinfo{author}{\bibfnamefont{J.}~\bibnamefont{Tailleur}} \bibnamefont{and}
  \bibinfo{author}{\bibfnamefont{M.~E.} \bibnamefont{Cates}},
  \bibinfo{journal}{Phys. Rev. Lett.} \textbf{\bibinfo{volume}{100}},
  \bibinfo{pages}{218103} (\bibinfo{year}{2008}).

\bibitem[{\citenamefont{Cates and Tailleur}(2013)}]{cate13}
\bibinfo{author}{\bibfnamefont{M.~E.} \bibnamefont{Cates}} \bibnamefont{and}
  \bibinfo{author}{\bibfnamefont{J.}~\bibnamefont{Tailleur}},
  \bibinfo{journal}{EPL} \textbf{\bibinfo{volume}{101}}, \bibinfo{pages}{20010}
  (\bibinfo{year}{2013}).

\bibitem[{\citenamefont{Ni et~al.}(2013)\citenamefont{Ni, Stuart, and
  Dijkstra}}]{ran13}
\bibinfo{author}{\bibfnamefont{R.}~\bibnamefont{Ni}},
  \bibinfo{author}{\bibfnamefont{M.~A.~C.} \bibnamefont{Stuart}},
  \bibnamefont{and} \bibinfo{author}{\bibfnamefont{M.}~\bibnamefont{Dijkstra}},
  \bibinfo{journal}{Nat. Commun.} \textbf{\bibinfo{volume}{4}},
  \bibinfo{pages}{2704} (\bibinfo{year}{2013}).

\bibitem[{\citenamefont{Berthier}(2013)}]{bert13b}
\bibinfo{author}{\bibfnamefont{L.}~\bibnamefont{Berthier}},
  \bibinfo{journal}{arXiv:1307.0704}  (\bibinfo{year}{2013}).

\bibitem[{\citenamefont{Bialk\'e et~al.}(2012)\citenamefont{Bialk\'e, Speck,
  and L\"owen}}]{bial12}
\bibinfo{author}{\bibfnamefont{J.}~\bibnamefont{Bialk\'e}},
  \bibinfo{author}{\bibfnamefont{T.}~\bibnamefont{Speck}}, \bibnamefont{and}
  \bibinfo{author}{\bibfnamefont{H.}~\bibnamefont{L\"owen}},
  \bibinfo{journal}{Phys. Rev. Lett.} \textbf{\bibinfo{volume}{108}},
  \bibinfo{pages}{168301} (\bibinfo{year}{2012}).

\bibitem[{\citenamefont{Menzel and L\"owen}(2013)}]{menz13}
\bibinfo{author}{\bibfnamefont{A.~M.} \bibnamefont{Menzel}} \bibnamefont{and}
  \bibinfo{author}{\bibfnamefont{H.}~\bibnamefont{L\"owen}},
  \bibinfo{journal}{Phys. Rev. Lett.} \textbf{\bibinfo{volume}{110}},
  \bibinfo{pages}{055702} (\bibinfo{year}{2013}).

\bibitem[{\citenamefont{Schwarz-Linek et~al.}(2012)\citenamefont{Schwarz-Linek,
  Valeriani, Cacciuto, Cates, Marenduzzo, Morozov, and Poon}}]{line12}
\bibinfo{author}{\bibfnamefont{J.}~\bibnamefont{Schwarz-Linek}},
  \bibinfo{author}{\bibfnamefont{C.}~\bibnamefont{Valeriani}},
  \bibinfo{author}{\bibfnamefont{A.}~\bibnamefont{Cacciuto}},
  \bibinfo{author}{\bibfnamefont{M.~E.} \bibnamefont{Cates}},
  \bibinfo{author}{\bibfnamefont{D.}~\bibnamefont{Marenduzzo}},
  \bibinfo{author}{\bibfnamefont{A.~N.} \bibnamefont{Morozov}},
  \bibnamefont{and} \bibinfo{author}{\bibfnamefont{W.~C.~K.}
  \bibnamefont{Poon}}, \bibinfo{journal}{Proc. Natl. Acad. Sci. U.S.A.}
  \textbf{\bibinfo{volume}{109}}, \bibinfo{pages}{4052} (\bibinfo{year}{2012}).

\bibitem[{\citenamefont{Redner et~al.}(2013{\natexlab{b}})\citenamefont{Redner,
  Baskaran, and Hagan}}]{redn13a}
\bibinfo{author}{\bibfnamefont{G.~S.} \bibnamefont{Redner}},
  \bibinfo{author}{\bibfnamefont{A.}~\bibnamefont{Baskaran}}, \bibnamefont{and}
  \bibinfo{author}{\bibfnamefont{M.~F.} \bibnamefont{Hagan}},
  \bibinfo{journal}{Phys. Rev. E} \textbf{\bibinfo{volume}{88}},
  \bibinfo{pages}{012305} (\bibinfo{year}{2013}{\natexlab{b}}).

\bibitem[{\citenamefont{Mognetti et~al.}(2013)\citenamefont{Mognetti, \ifmmode
  \check{S}\else \v{S}\fi{}ari\ifmmode~\acute{c}\else \'{c}\fi{},
  Angioletti-Uberti, Cacciuto, Valeriani, and Frenkel}}]{mogn13}
\bibinfo{author}{\bibfnamefont{B.~M.} \bibnamefont{Mognetti}},
  \bibinfo{author}{\bibfnamefont{A.}~\bibnamefont{\ifmmode \check{S}\else
  \v{S}\fi{}ari\ifmmode~\acute{c}\else \'{c}\fi{}}},
  \bibinfo{author}{\bibfnamefont{S.}~\bibnamefont{Angioletti-Uberti}},
  \bibinfo{author}{\bibfnamefont{A.}~\bibnamefont{Cacciuto}},
  \bibinfo{author}{\bibfnamefont{C.}~\bibnamefont{Valeriani}},
  \bibnamefont{and} \bibinfo{author}{\bibfnamefont{D.}~\bibnamefont{Frenkel}},
  \bibinfo{journal}{Phys. Rev. Lett.} \textbf{\bibinfo{volume}{111}},
  \bibinfo{pages}{245702} (\bibinfo{year}{2013}).

\bibitem[{\citenamefont{Wittkowski et~al.}(2013)\citenamefont{Wittkowski,
  Tiribocchi, Stenhammar, Allen, Marenduzzo, and Cates}}]{witt13a}
\bibinfo{author}{\bibfnamefont{R.}~\bibnamefont{Wittkowski}},
  \bibinfo{author}{\bibfnamefont{A.}~\bibnamefont{Tiribocchi}},
  \bibinfo{author}{\bibfnamefont{J.}~\bibnamefont{Stenhammar}},
  \bibinfo{author}{\bibfnamefont{R.~J.} \bibnamefont{Allen}},
  \bibinfo{author}{\bibfnamefont{D.}~\bibnamefont{Marenduzzo}},
  \bibnamefont{and} \bibinfo{author}{\bibfnamefont{M.~E.} \bibnamefont{Cates}},
  \bibinfo{journal}{arXiv:1311.1256}  (\bibinfo{year}{2013}).

\bibitem[{\citenamefont{Cates et~al.}(2010)\citenamefont{Cates, Marenduzzo,
  Pagonabarraga, and Tailleur}}]{cate10}
\bibinfo{author}{\bibfnamefont{M.~E.} \bibnamefont{Cates}},
  \bibinfo{author}{\bibfnamefont{D.}~\bibnamefont{Marenduzzo}},
  \bibinfo{author}{\bibfnamefont{I.}~\bibnamefont{Pagonabarraga}},
  \bibnamefont{and} \bibinfo{author}{\bibfnamefont{J.}~\bibnamefont{Tailleur}},
  \bibinfo{journal}{Proc. Natl. Acad. Sci. U.S.A.}
  \textbf{\bibinfo{volume}{107}}, \bibinfo{pages}{11715}
  (\bibinfo{year}{2010}).

\bibitem[{\citenamefont{Cross and Hohenberg}(1993)}]{cros93}
\bibinfo{author}{\bibfnamefont{M.~C.} \bibnamefont{Cross}} \bibnamefont{and}
  \bibinfo{author}{\bibfnamefont{P.~C.} \bibnamefont{Hohenberg}},
  \bibinfo{journal}{Rev. Mod. Phys.} \textbf{\bibinfo{volume}{65}},
  \bibinfo{pages}{851} (\bibinfo{year}{1993}).

\bibitem[{\citenamefont{Cahn and Hilliard}(1958)}]{cahn58}
\bibinfo{author}{\bibfnamefont{J.~W.} \bibnamefont{Cahn}} \bibnamefont{and}
  \bibinfo{author}{\bibfnamefont{J.~E.} \bibnamefont{Hilliard}},
  \bibinfo{journal}{J. Chem. Phys.} \textbf{\bibinfo{volume}{28}},
  \bibinfo{pages}{258} (\bibinfo{year}{1958}).

\bibitem[{\citenamefont{Novick-Cohen}(1985)}]{novi85}
\bibinfo{author}{\bibfnamefont{A.}~\bibnamefont{Novick-Cohen}},
  \bibinfo{journal}{J. Stat. P} \textbf{\bibinfo{volume}{38}},
  \bibinfo{pages}{707} (\bibinfo{year}{1985}).

\end{thebibliography}
\end{document}